\documentstyle[procl]{article}
\def\Journal#1#2#3#4{{#1} {\bf #2}, #3 (#4)}

\def\PR{\em Phys. Rep.}

\def\NPB{{\em Nucl. Phys.} B}

\def\PLB{{\em Phys. Lett.}  B}

\def\PRD{{\em Phys. Rev.} D}

\newcommand{\bfk}{\mbox{\boldmath $k$}}
\newcommand{\bfp}{\mbox{\boldmath $p$}}
\def\nostrocostruttino#1\over#2{\mathrel{\mathop{\kern 0pt \rlap
{\hbox{$#1$}}} \hbox{\kern-.135em $#2$}}}
\def\sumint{\nostrocostruttino \sum \over {\displaystyle\int}}

\begin{document}

\vbox{ \null\hfill DFTT 33/96 \\
\null\hfill INFNCA-TH9615 \\
\null\hfill hep-ph/9607334}

\vspace{0.5truecm}

\title{INCLUSIVE PRODUCTION OF HADRONS IN
$\ell^\uparrow p^\uparrow \to h^\uparrow X$ 
AND SPIN MEASUREMENTS$\,$\footnote{ $\,$Talk delivered by M. Anselmino at
DIS96, Rome, April 15-19, 1996.}}
\setcounter{footnote}{0}
\renewcommand{\thefootnote}{\fnsymbol{footnote}}

\author{M. ANSELMINO, M. BOGLIONE}

\address{Dipartimento di Fisica Teorica, Universit\`a di Torino and \\
      INFN, Sezione di Torino, Via P. Giuria 1, I-10125 Torino, Italy}

\author{J. HANSSON}

\address{Department of Physics, Lule{\aa} University of Technology, S-97187
      Lule\aa, Sweden}

\author{F. MURGIA}

\address{INFN, Sezione di Cagliari, Via A. Negri 18, I-09127 Cagliari, Italy} 

\maketitle\abstracts{
We discuss the production of polarized hadrons in polarized lepton
nucleon interactions and show that the helicity density matrix 
of the hadron, when measurable, can give information on the spin 
structure of the nucleon and the spin dependence of the quark 
fragmentation process. Single spin asymmetries in the $\ell N^\uparrow 
\to hX$ process are also briefly discussed.}

According to the QCD hard scattering scheme and the factorization theorem
\cite{col1}$^{\!-\,}$\cite{col3}$^{\!,\,}$\cite{old}$^{\!,\,}$\cite{noi} 
the helicity density matrix 
of the hadron $h$ inclusively produced in the DIS process 
$\ell^\uparrow N^\uparrow \to h^\uparrow X$ is given by 
\begin{eqnarray}
\rho_{\lambda^{\,}_h,\lambda^\prime_h}^{(s,S)}(h) &\>& \!\!\!\!\!\!\!\!\!
\frac{E_h \, d^3\sigma^{\ell,s + N,S \to h + X}} {d^{3} \bfp_h} = 
\sum_{q; \lambda^{\,}_{\ell}, \lambda^{\,}_q, \lambda^\prime_q}
\int \frac {dx}{\pi z} \frac {1}{16 \pi x^2 s^2} \label{gen} \\  
& & \rho^{\ell,s}_{\lambda^{\,}_{\ell}, \lambda^{\,}_{\ell}} \,
\rho_{\lambda^{\,}_q, \lambda^{\prime}_q}^{q/N,S} \, f_{q/N}(x) \,
\hat M^q_{\lambda^{\,}_{\ell}, \lambda^{\,}_q; 
\lambda^{\,}_{\ell}, \lambda^{\,}_q} \, 
\hat M^{q*}_{\lambda^{\,}_{\ell}, \lambda^{\prime}_q; 
\lambda^{\,}_{\ell}, \lambda^{\prime}_q} \,
D_{\lambda^{\,}_h, \lambda^{\prime}_h}^{\lambda^{\,}_q,\lambda^{\prime}_q}(z) 
\nonumber
\end{eqnarray}
where $\rho^{\ell,s}$ is the helicity density matrix of the initial 
lepton with spin $s$, $f_{q/N}(x)$ is the number density of unpolarized 
quarks $q$ with momentum fraction $x$ inside an unpolarized nucleon and 
$\rho^{q/N,S}$ is the helicity density matrix of quark $q$ inside the 
polarized nucleon $N$ with spin $S$. 
The $\hat M^q_{\lambda^{\,}_{\ell}, \lambda^{\,}_q; \lambda^{\,}_{\ell}, 
\lambda^{\,}_q}$'s are the helicity amplitudes for the elementary process 
$\ell q \to \ell q$. The final lepton spin is not observed and helicity 
conservation of perturbative QCD and QED has already been taken into account
in the above equation: as a consequence only the diagonal elements of 
$\rho^{\ell,s}$ contribute to $\rho(h)$ and non diagonal elements, present in 
case of transversely polarized leptons, do not contribute.
$D^{\lambda^{\,}_q,\lambda^{\prime}_q}_{\lambda^{\,}_h,\lambda^{\prime}_h}(z)$ 
is the product of {\it fragmentation amplitudes}
\begin{equation}
D^{\lambda^{\,}_q,\lambda^{\prime}_q}_{\lambda^{\,}_h,\lambda^{\prime}_h}(z) 
=\sumint_{X,\lambda^{\,}_X} {\cal D}_{\lambda^{\,}_{X}, \lambda_h; 
\lambda^{\,}_q} \,
{\cal D}^*_{\lambda^{\,}_{X}, \lambda^{\prime}_h; \lambda^{\prime}_q}
\label{framp}
\end{equation}
where the $\sumint_{X,\lambda^{\,}_X}$ stands for a spin sum and phase space
integration of the undetected particles, considered as a system $X$.
The usual unpolarized fragmentation function $D_{h/q}(z)$, {\it i.e.} 
the density number of hadrons $h$ resulting from the fragmentation of 
an unpolarized quark $q$ and carrying a fraction $z$ of its momentum,
is given by
\begin{equation}
D_{h/q}(z) = {1\over 2} \sum_{\lambda^{\,}_q, \lambda^{\,}_h} 
D^{\lambda^{\,}_q,\lambda^{\,}_q}_{\lambda^{\,}_h,\lambda^{\,}_h}(z) 
= {1\over 2} \sum_{\lambda^{\,}_q, \lambda^{\,}_h} 
D_{h_{\lambda^{\,}_h}/q_{\lambda^{\,}_q}}(z) \,,
\label{fr}
\end{equation}
where $D^{\lambda^{\,}_q,\lambda^{\,}_q}_{\lambda^{\,}_h,
\lambda^{\,}_h}(z) \equiv D_{h_{\lambda^{\,}_h}/q_{\lambda^{\,}_q}}$ is
a polarized fragmentation function, {\it i.e.} the density number of 
hadrons $h$ with helicity $\lambda^{\,}_h$ resulting from the fragmentation 
of a quark $q$  with helicity $\lambda^{\,}_q$. Notice that
by definition and parity invariance the generalized fragmentation 
functions (\ref{framp}) obey the relationships
\begin{eqnarray}
D^{\lambda^{\,}_q,\lambda^{\prime}_q}_{\lambda^{\,}_h,\lambda^{\prime}_h}
&=& \left( D^{\lambda^{\prime}_q,\lambda^{\,}_q}
_{\lambda^{\prime}_h,\lambda^{\,}_h} \right)^* \, \label{def} \\
D^{-\lambda^{\,}_q,-\lambda^{\prime}_q}_{-\lambda^{\,}_h,-\lambda^{\prime}_h}
&=& -(-1)^{2S_h}(-1)^{\lambda^{\,}_q + \lambda^{\prime}_q
+ \lambda^{\,}_h + \lambda^{\prime}_h} \>
D^{\lambda^{\,}_q,\lambda^{\prime}_q}_{\lambda^{\,}_h,\lambda^{\prime}_h} \,,
\label{par}
\end{eqnarray}
where $S_h$ is the hadron spin; notice also that collinear configuration 
(intrinsic $\bfk_{\perp} =0$) together with angular 
momentum conservation in the forward fragmentation process imply
\begin{equation}
D^{\lambda^{\,}_q,\lambda^{\prime}_q}_{\lambda^{\,}_h,\lambda^{\prime}_h}
= 0 \quad\quad {\mbox{\rm when}} \quad\quad
\lambda^{\,}_q - \lambda^{\prime}_q \not= \lambda^{\,}_h-\lambda^{\prime}_h \,.
\label{for}
\end{equation}

Eq. (\ref{gen}) holds at leading twist, leading order in the 
coupling constants and large $Q^2$ values; the intrinsic $\bfk_\perp$ 
of the partons have been integrated over and collinear configurations 
dominate both the distribution functions and the fragmentation processes. 
For simplicity of notations we have not indicated the $Q^2$ scale 
dependences in $f$ and $D$; the variable $z$ is related to 
$x$ by the usual imposition of energy momentum conservation
in the elementary 2 $\to$ 2 process; more technical details can be 
found in Ref. \cite{noi}. 

The quark helicity density matrix $\rho^{q/N,S}$ can be decomposed as
\begin{equation}
\rho^{q/N,S} = P_P^{q/N,S} \rho^{N,S} + P_A^{q/N,S} \rho^{N,-S}
\label{rhoq}
\end{equation}
where $P_{P(A)}^{q/N,S}$ (which, in general, depends on $x$)
is the probability that the spin of the quark 
inside the polarized nucleon $N$ is parallel (antiparallel) to the nucleon 
spin $S$ and $\rho^{N,S(-S)}$ is the helicity density matrix of the nucleon 
with spin $S(-S)$. Notice that 
\begin{equation}
P^{q/N,S} = P_P^{q/N,S} - P_A^{q/N,S}  
\label{polq}
\end{equation}
is the component of the quark polarization vector along the parent nucleon 
spin direction. In more familiar notations one has, for longitudinally
polarized nucleons
\begin{equation}
f_{q/N}(x) \, P_{P(A)}^{q/N,S_L}(x) = f_{q_{+(-)}/N_+}(x) = f_{q_{-(+)}/N_-}(x)
\label{pold}
\end{equation}
where $f_{q_{+(-)}/N_+}$ is the polarized distribution function, that is 
the density number of quarks with helicity $+(-)$ inside a nucleon with 
helicity + and the last equality holds due to parity invariance. 
This implies 
\begin{equation}
f_{q/N} \, P^{q/N,S_L} = f_{q_+/N_+} - f_{q_-/N_+} \equiv \Delta q 
\label{dq}
\end{equation}
and similarly for transverse polarization $T$,
\begin{equation}
f_{q/N} \, P^{q/N,S_T} = f_{q,S_T/N,S_T} - f_{q,-S_T/N,S_T} 
\equiv \Delta_T q \,.
\label{dtq}
\end{equation}
where $P^{q/N,S_T}$ is the quark transverse polarization in a transversely 
polarized nucleon. 

We shall now consider several particular cases of Eq. (\ref{gen}) and 
discuss what can be learned or expected from a measurement of $\rho(h)$. 
We consider both the case of spin 1 and spin 1/2 hadrons $h$, at leading 
twist only; higher twist contribution to single spin asymmetries in
$\ell N^\uparrow \to hX$ processes will be shortly discussed at the end. 
Somewhat similar analyses can be found in Ref. \cite{mul} and \cite{jm}.

We choose $xz$ as the hadron production plane with the lepton moving along 
the $z$-axis and the nucleon in the opposite direction in the lepton-nucleon
centre of mass frame; as usual we indicate 
by an index $L$ the (longitudinal) nucleon spin orientation along the 
$z$-axis, by an index $S$ the (sideway) orientation along the $x$-axis and 
by an index $N$ the (normal) orientation along the $y$-axis.

\goodbreak

\vskip 6pt
\noindent
{\it Spin 1 final hadron; unpolarized leptons and longitudinally polarized 
nucleons}
\vskip 6pt
\nobreak
In this case Eqs. (\ref{gen}) and (\ref{for}) yield (using obvious 
shorter notations)
\begin{eqnarray}
\rho_{1,1}^{(S_L)}(V) \, d^3\sigma &=& 
\sum_q \int \frac {dx}{\pi z} \, f_{q/N} \, d\hat\sigma^q
\left[ P_A^{q/N,S_L} D_{V_1/q_+}
     + P_P^{q/N,S_L} D_{V_1/q_-} \right] \label{vl11} \nonumber \\
\rho_{0,0}^{(S_L)}(V) \, d^3\sigma &=& 
\sum_q \int \frac {dx}{\pi z} \, f_{q/N} \, d\hat\sigma^q \, 
D_{V_0/q_+} \label{vl00} \\
\rho_{-1,-1}^{(S_L)}(V) \, d^3\sigma &=& 
\sum_q \int \frac {dx}{\pi z} \, f_{q/N} \, d\hat\sigma^q
\left[ P_A^{q/N,S_L} D_{V_1/q_-}
     + P_P^{q/N,S_L} D_{V_1/q_+} \right] \label{vl-1-1} \nonumber
\end{eqnarray}
where the apex $(S_L)$ reminds of the nucleon spin configuration.

\vskip 6pt
\noindent
{\it Spin 1 final hadron; unpolarized leptons and transversely polarized 
nucleons, T=S,N}
\vskip 6pt

In this case we have both diagonal 
\begin{eqnarray}
\rho_{1,1}^{(S_T)}(V) \, d^3\sigma &=& 
\sum_q \int \frac {dx}{\pi z} \, f_{q/N} \, d\hat\sigma^q \,
{1 \over 2} \left[ D_{V_1/q_+} + D_{V_1/q_-} \right] \label{vs11} \\
\rho_{0,0}^{(S_T)}(V) \, d^3\sigma &=& 
\sum_q \int \frac {dx}{\pi z} \, f_{q/N} \, d\hat\sigma^q \,
D_{V_0/q_+} \label{vs00} \\
\rho_{-1,-1}^{(S_T)}(V) &=& \rho_{1,1}^{(S_T)}(V) \label{vs-1-1}
= {1 - \rho_{0,0}^{(S_T)}(V) \over 2}
\end{eqnarray}
and non diagonal matrix elements
\begin{eqnarray}
\rho_{1,0}^{(S_S)}(V) \, d^3\sigma &=& 
\sum_q \int \frac {dx}{\pi z} \, f_{q/N} \, {P^{q/N,S_S} \over 2} 
\left[ \mbox{\rm Re} \hat M_+^q \hat M_-^{q*} \right]
D_{1,0}^{+,-} \label{vs10} \\
\rho_{-1,0}^{(S_S)}(V) &=& \rho_{1,0}^{(S_S)}(V) \label{vs-10} \\
\rho_{1,0}^{(S_N)}(V) &=& - \rho_{-1,0}^{(S_N)}(V) 
= i \rho_{1,0}^{(S_S)}(V) \,. \label{vn-10} 
\end{eqnarray}
which involve the non diagonal fragmentation functions (\ref{framp}); 
$\hat M_{+(-)}$ is a short notation for 
$\hat M_{+,+;+,+(+,-;+,-)}/4\sqrt {\hat s}$.

\vskip 6pt
\noindent
{\it Spin 1/2 final hadron; unpolarized leptons and polarized nucleons}
\vskip 6pt

In case of final spin 1/2 hadrons ($h=B$), with unpolarized leptons and 
spin $S$ nucleons, we have the non zero results
\begin{eqnarray}
P_x^{(S_S)} \, d^3\sigma &=& 
\sum_q \int \frac {dx}{\pi z} \, f_{q/N} \, P^{q/N,S_S} 
\left[ \mbox{\rm Re} \hat M_+^q \hat M_-^{q*} \right]
D_{+,-}^{+,-} \nonumber \\
&=& \sum_q \int \frac {dx}{\pi z} \, \Delta_T q \> \Delta_N \hat\sigma^q \>
\Delta_T D_{B/q} \label{px} \\
P_y^{(S_N)} &=& - P_x^{(S_S)} \label{py} \\
P_z^{(S_L)} \, d^3\sigma &=&
\sum_q \int \frac {dx}{\pi z} \, f_{q/N} \, P^{q/N,S_L} \, 
d\hat\sigma^q \, \left[ D_{B_+/q_-} - D_{B_+/q_+} \right] \nonumber \\
&-& \sum_q \int \frac {dx}{\pi z} \, \Delta q \> d\hat\sigma^q \> 
\Delta D_{B/q} \,. \label{pz}
\end{eqnarray}
where $P_i =$ Tr$(\sigma^i\rho)$ are the components of the polarization 
vector in the helicity rest frame of hadron $B$, and where the apices 
$S_L, S_N$ and $S_S$ refer to the nucleon spin orientations in the reference 
frame where we compute the scattering. In the above equations
\begin{equation}
- \left[ \mbox{\rm Re} \hat M_+^q \hat M_-^{q*} \right]
= {d\hat\sigma^{\ell + q,S_N \to \ell + q,S_N} \over d\hat t}
- {d\hat\sigma^{\ell + q,S_N \to \ell + q,-S_N}\over d\hat t}
\equiv \Delta_N \hat\sigma^q \,,
\label{nbas}
\end{equation}
\begin{equation}
D_{+,-}^{+,-} = D_{B,S_N/q,S_N} - D_{B,-S_N/q,S_N}
\equiv \Delta_T D_{B/q} \,,
\label{dtd}
\end{equation}
which is a difference of transverse fragmentation functions and 
\begin{equation}
D_{B,S_L/q,S_L} - D_{B,-S_L/q,S_L} = D_{B_+/q_+} - D_{B_-/q_+}
\equiv \Delta D_{B/q} \,.
\label{dd}
\end{equation}

\vskip 6pt
\noindent
{\it Longitudinally polarized leptons and polarized nucleons}
\vskip 6pt

We discuss now the case of polarized leptons; as we noticed after 
Eq. (\ref{gen}) only the diagonal elements of the lepton helicity density 
matrix $\rho^{\ell,s}$ contribute to $\rho(h)$, so that only longitudinal 
polarizations could affect the results. We consider then longitudinally 
polarized leptons, $s=s_L$.

One obtains the same results as in the unpolarized lepton case for the non 
diagonal matrix elements and slightly different ones for the diagonal 
elements, for example
\begin{eqnarray}
\!\!\! \rho_{0,0}^{(s_L,S_L)}(V) \, d^3\sigma_L \!\!\!&=&\!\!\!
\sum_q \int \frac {dx}{\pi z}
\left[ f_{q_-/N_+} \, |\hat M_+^q|^2 
     + f_{q_+/N_+} \, |\hat M_-^q|^2 \right] \, D_{V_0/q_+}
\label{vl00pl} \\
\!\!\! P_z^{(s_L,S_T)}(B) \, d^3\sigma \!\!\!&=&\!\!\!
\sum_q \int \frac {dx}{\pi z} \, f_{q/N} \, {1 \over 2} 
\left[ |\hat M_+^q|^2 -  |\hat M_-^q|^2 \right] \Delta D_{B/q} \,. 
\label{pztpl}
\end{eqnarray}

\goodbreak

\vskip 6pt
\noindent
{\it Experimental measurements} 
\vskip 6pt
\nobreak
Some elements of the helicity density matrix of the produced hadrons can be 
measured via the angular distribution of the final hadron $h$ decay; 
typical examples are the $\rho \to \pi\pi$ and $\Lambda \to p \pi$ decays.

For spin 1 hadrons one can measure $\rho_{0,0}$ and $\rho_{1,0}$ 
[Eqs. (\ref{vl00}) and (\ref{vs10})], whereas for weakly 
decaying spin 1/2 hadrons one measures $P_y^{S_N}$ and $P_z^{S_L}$
[Eqs. (\ref{py}) and (\ref{pz})]. Such measurements
supply information on the polarized quark fragmentation process
and the polarized distribution functions. Further discussion and 
an estimate of $\rho_{1,0}$ can be found in Ref. \cite{noi}. We 
only remind here that according to $SU(6)$ wave function the entire 
$\Lambda$ polarization is 
due to the strange quark, so that the difference of polarized fragmentation
functions in Eq. (\ref{pz}) is different from zero only for $s$ quarks,
$\Delta D_{\Lambda/s} = D_{\Lambda/s}$. Then 
Eq. (\ref{pz}) reads
\begin{equation}
P_z^{(S_L)} = -\frac{\int dx \, (\pi z)^{-1} \, \Delta s \, 
d\hat\sigma^s \, D_{\Lambda/s}}
{\sum_q \int dx \, (\pi z)^{-1} \, f_{q/N} \, d\hat\sigma^q \, D_{\Lambda/q}}
\label{pzla}
\end{equation}
Such a quantity is expected to be rather small; however, any non zero value
would offer valuable information on the much debated issue of longitudinal 
strange quark polarization, $\Delta s$, inside a longitudinally polarized 
nucleon. A similar information on the transverse polarization can 
be obtained from a measurement of $P_y^{(S_N)}$ and Eq. (\ref{py}).

\vskip 6pt
\noindent
{\it Single spin asymmetries and $\bfk_\perp$ effects in $\ell N^\uparrow 
\to hX$} 
\vskip 6pt

We conclude by mentioning that a fundamental property of quark fragmentation, 
the quark analysing power
\begin{equation}
A_{h/q} = \frac{\tilde D_{h/q^\uparrow}(z, \bfk_\perp)
- \tilde D_{h/q^\downarrow}(z, \bfk_\perp)} 
{\tilde D_{h/q^\uparrow}(z, \bfk_\perp)
+ \tilde D_{h/q^\downarrow}(z, \bfk_\perp)}
=\frac{\tilde D_{h/q^\uparrow}(z, \bfk_\perp)
- \tilde D_{h/q^\uparrow}(z, -\bfk_\perp)} 
{\tilde D_{h/q^\uparrow}(z, \bfk_\perp)
+ \tilde D_{h/q^\uparrow}(z, -\bfk_\perp)}
\label{ce}
\end{equation}
has been recently proposed \cite{col2}$^{\!,\,}$\cite{col3} and suggested to 
be sizeable. $\tilde D_{h/q^\uparrow}(z, \bfk_\perp)$ denotes the $\bfk_\perp$
dependent fragmentation function of the polarized quark $q$.

Such an effect, a leading twist one, might be measured by looking at the 
production of, say, pions with opposite intrinsic $\bfk_\perp$ inside the 
current jet in the scattering of unpolarized leptons off transversely 
polarized protons; or, equivalently, to the production of pions with a 
certain $\bfk_\perp$ resulting from up or down transversely polarized protons.

This same effects might also be responsible, as suggested for the 
process $p^\uparrow p \to \pi X$ \cite{art}$^{\!-\,}$\cite{tri}, for the higher 
twist single spin asymmetry 
\begin{equation}
A_N = \frac
{d\sigma^{\ell N^\uparrow \to hX} - d\sigma^{\ell N^\downarrow \to hX}} 
{d\sigma^{\ell N^\uparrow \to hX} + d\sigma^{\ell N^\downarrow \to hX}}
\end{equation}  
where integration is performed over the hadron $h$ intrinsic $\bfk_\perp$.
A measurement of such an asymmetry would provide valuable information.

\end{document}